\documentclass{iopconfser}

\usepackage{a4wide,amssymb,cite,graphicx}
\usepackage{epsfig}
\usepackage {mathtools}
\usepackage{subfig}
\usepackage{booktabs}  
\usepackage{threeparttable}
\usepackage{caption}
\usepackage{float}

\begin{document}

\noindent Conference Proceedings for BCVSPIN 2024: Particle Physics and Cosmology in the Himalayas\\Kathmandu, Nepal, December 9-13, 2024 

\title{Exploring new physics in the FCNC $B_c \rightarrow D_s\nu\bar{\nu}$ channel}

\author{Tarun Kumar$^{1}$ and Barilang Mawlong$^{2}$}

\affil{$^{1,2}$School of Physics, University of Hyderabad, Hyderabad-500046, India}


\email{tarunsharma8501@gmail.com$^1$, barilang05@gmail.com$^2$ }

\begin{abstract}
The rare $b \to s \ell^+\ell^-$ transition is a well-explored transition, particularly for new physics beyond the standard model. Similar to this transition, another flavor changing neutral current transition $b \to s \nu\bar{\nu}$ involving a dineutrino pair also plays an important role in the search for new physics. The $B \to K \nu\bar{\nu}$ mode is one such dineutrino channel that has been analyzed in many works both within the standard model and beyond standard model framework. In this paper, we investigate the $B_c \to D_s \nu\bar{\nu}$ decay channel which proceeds via the $b \to s \nu\bar{\nu}$ transition. We consider an effective Hamiltonian framework involving new physics currents. To explore these currents further, we also analyse the decay mode within the $V_2$ leptoquark model. In our analysis, we use the form factors that have been obtained in lattice QCD calculations. We then make predictions of the $B_c \to D_s \nu\bar{\nu}$ branching fraction both within the standard model and beyond.
\end{abstract}

\section{Introduction}

In flavor changing neutral current (FCNC) transitions, such as the $b \to s 
\ell \bar{\ell}$ transitions, anomalies were observed in the measured and predicted values of the lepton flavor universality ratio $R_{K^{(\ast)}} = Br(B \to K^{(\ast)} \mu^+ \mu^-)/Br(B \to K^{(\ast)} e^+ e^-)$. The recent measured values of $R_{K^{(\ast)}}$ as reported by LHCb \cite{LHCb:2022vje} and CMS  \cite{CMS:2024syx} collaborations are compatible with SM predictions at $1 \sigma$. However, the measured branching fraction $Br(B^+ \to K^+ \mu^+ \mu^-)$ in the central-$q^2$ region ($1.1< q^2 < 6.0$ GeV$^2$) is observed to have a tension with the $SM$ value at $\sim 4 \sigma$ \cite{Parrott:2022zte}. Similarly, in the dineutrino $b \to s \nu\Bar{\nu} $ transition, the measured branching fraction $Br(B \to K \nu\Bar{\nu})$ is observed to be above the SM prediction at about $2.7\sigma$ \cite{Belle-II:2023esi}. 
Thus, the possibility of NP in FCNC transitions cannot be neglected. For our analysis, we focus on the FCNC $B_c \to D_s \nu\Bar{\nu}$ channel. We use the available lattice QCD (LQCD) form factors \cite{Cooper:2021bkt} for the $B_c \to D_s$ transition in our calculations. We analyse this mode using a model independent (MI) approach and within the $V^{\mu}_2$ leptoquark (LQ) framework. We obtain the required NP couplings for both frameworks. For the MI approach, the couplings are constrained using the measured branching fraction $Br(B \to K \nu \bar{\nu})$. For the LQ scenario, we constrain the couplings using the branching fractions $Br (B \to K \ell \bar{\ell})$ and $Br (B \to K \nu \bar{\nu})$. In this work, we consider the Wilson coefficients to be lepton flavor universal. 


\section{Theoretical Framework}
The effective Hamiltonian for $b \to s \nu \Bar{\nu} $ transitions can be written as \cite{Altmannshofer:2009ma}
\begin{align}
\mathcal{H}_{\text{eff}} = &  -\frac{4 G_F}{\sqrt{2}} V_{tb} V_{ts}^* \frac{\alpha_{EW}}{4 \pi} 
\Big[ C_L^\nu (\bar{s} \gamma_\mu P_L b) (\bar{\nu} \gamma^\mu (1 - \gamma_5) \nu) + C_R^\nu (\bar{s} \gamma_\mu P_R b) (\bar{\nu} \gamma^\mu (1 - \gamma_5) \nu) \Big],
\end{align}
where $C_L^{\nu} \text{~and~} C_R^\nu$ are the Wilson coefficients (WCs), and $ C_L^\nu$ can be written as $(C_L^{\nu~SM} + C_L^{\nu~NP})$. \\

The differential branching fraction for $B \to P \nu\Bar{\nu}$ ($P$ is a pseudoscalar) decays is given by
\begin{equation*}
\frac{dBr}{dq^2} = \tau_{B} 3 |N|^2 \rho_{P}(q^2)|C_L^{\nu} + C_R^{\nu}|^2,
\end{equation*}
where \begin{equation}
\begin{aligned}
    N = V_{tb} V_{ts}^* \frac{G_F \alpha_{EW}}{16 \pi^2} \sqrt{\frac{m_{B}}{3\pi}} {~\text{~~and}~} \quad &
   \rho_P(q^2) = \frac{\lambda_P^{3/2}(q^2)}{m_B^4} \left[ f_+^P(q^2) \right]^2,
\end{aligned}
\end{equation} and $ f_+^P(q^2)$ is the $B \to P$ form factor. Other details can be found in \cite{Buras:2014fpa}. \\

Including NP contributions, the effective Hamiltonian for $b \to s \ell \Bar{\ell} $ transitions can be written as \cite{Buras:1994dj, Capdevila:2017bsm} 
\begin{align}
\mathcal{H}_{\text{eff}} = & -\frac{\alpha_{EW} G_F}{\sqrt{2} \pi} V_{tb} V_{ts}^* \Bigg[
     \frac{2 C_7^{\text{eff}}}{q^2} \big[ \bar{s} \sigma^{\mu\nu} q_\nu (m_s P_L + m_b P_R) b \big] (\bar{\ell} \gamma_\mu \ell) + [C_9^{\text{eff}} + C_9^{\text{NP}}](\bar{s} \gamma^\mu P_L b)(\bar{\ell} \gamma_\mu \ell)  \nonumber \\
    &
    + [C_{10} + C_{10}^{\text{NP}}] (\bar{s} \gamma^\mu P_L b)(\bar{\ell} \gamma_\mu \gamma_5 \ell) + C_{9}^{' \, \text{NP}}(\bar{s} \gamma^\mu P_R b)(\bar{\ell} \gamma_\mu \ell)
    + C_{10}^{' \, \text{NP}}(\bar{s} \gamma^\mu P_R b)(\bar{\ell} \gamma_\mu \gamma_5 \ell)
\Bigg],
\end{align}
where $C_7^{\rm{eff}}$, $C_9^{\text{eff}}, C_{9}^{(\prime)\text{NP}}, C_{10}^{ (\prime)\text{NP}}$ are the Wilson coefficients. \\



The differential branching fraction for $B \to P \ell\Bar{\ell}$ decays is given by 
\begin{equation}
\frac{dBr}{dq^2} = \tau_{B}(2a_\ell + \frac{2}{3}c_\ell),
\end{equation}
where
 \begin{eqnarray}
a_\ell &=& \frac{G_F^2 \alpha^2_{\text{EW}} |V_{tb} V_{ts}^*|^2}{2^9 \pi^5 m^3_{B}} 
\beta_\ell \sqrt{\lambda} \Bigg[ q^2 |F_P|^2 + \frac{\lambda}{4} \left( |F_A|^2 + |F_V|^2 \right) + 4m_\ell^2 m_{B}^2 |F_A|^2 \nonumber \\
&& + 2 m_\ell \left( m_{B}^2 - m_P^2 + q^2 \right) \text{Re}(F_P F_A^*) \Bigg], \nonumber \\
c_\ell &=& -\frac{G_F^2 \alpha^2_{\text{EW}} |V_{tb} V_{ts}^*|^2}{2^9 \pi^5 m^3_{B}} 
\beta_\ell \sqrt{\lambda} \frac{\lambda \beta_\ell^2}{4} 
\left( |F_A|^2 + |F_V|^2 \right),\nonumber
\end{eqnarray}In the above, $F_P$, $F_V$ and $F_A$ are functions of form factors and Wilson coefficients. Details can be found in \cite{Bouchard:2013eph}.



\section{Analysis}
\vspace{0.1cm}
%

\subsection{\textbf{Input parameters}}
\vspace{0.1cm}
Numerical values of various input parameters are taken from Particle Data Group \cite{ParticleDataGroup:2024cfk}. The values of the SM Wilson coefficients are taken from \cite{Altmannshofer:2008dz}. Form factors for the $B \to K$ transition are considered from fully relativistic lattice QCD \cite{Parrott:2022rgu} calculations.


\subsection{\textbf{Model Independent approach:}}
\vspace{0.1cm}
Here, the measured branching fraction of $(B \to K \nu\Bar{\nu})$ is used to constrain the NP WCs. We predict the range for these WCs as : $-9.20428 < (C_L^{\nu~NP} + C_R^{\nu})  < -5.00924$. Fig. \ref{fig:CL-CR} represents the allowed parametric space for the contributing WCs and Fig. \ref{fig:Bc-to-Ds} represents the variation of the differential branching fraction $dBr(B_c \to D_s \nu\Bar{\nu})$ as a function of $q^2$. The predicted branching fraction values for $B_c \to D_s \nu\Bar{\nu}$ within the SM and beyond SM are given in Table \ref{tab:bc-decays}.

\begin{minipage}{0.45\textwidth}
    \centering
    \includegraphics[width=\textwidth]{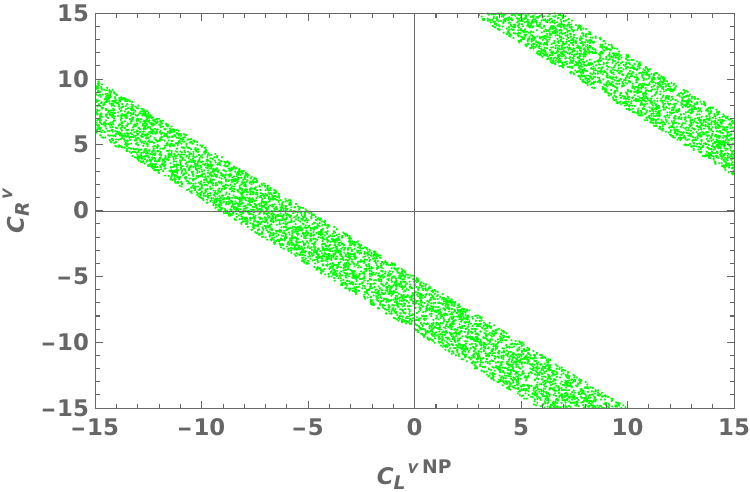} 
    \captionof{figure}{Constraint plot for $C_L^{\nu~NP}$ - $C_R^{\nu}$.}
    \label{fig:CL-CR}
\end{minipage}
\hfill
\begin{minipage}{0.45\textwidth}
    \centering
    \includegraphics[width=\textwidth]{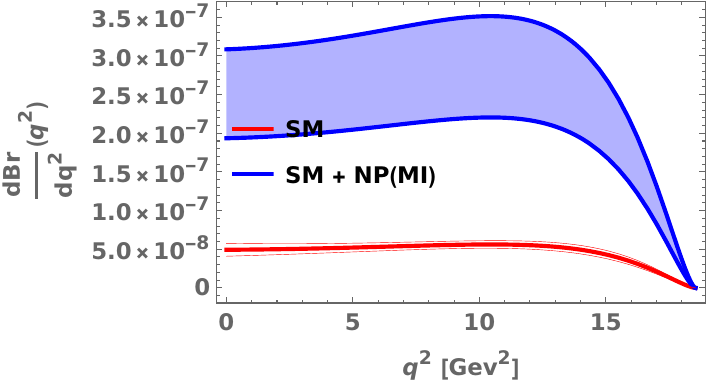} 
    \captionof{figure}{$dBr(B_c \to D_s \nu\Bar{\nu})$ as a function of $q^2$.}
    \label{fig:Bc-to-Ds}
\end{minipage}

\begin{table}[h]
    \centering
    \begin{tabular}{|c|c|}
        \hline
        $Br(B_c \to D_s \nu\Bar{\nu})_{SM}$ & $(8.70\pm 0.97) \times 10^{-7}$ \\
        \hline
        $Br(B_c \to D_s \nu\Bar{\nu})_{SM + NP(MI)}$(range) & $(3.1396 - 4.9239) \times 10^{-6}$\\
        \hline
    \end{tabular}
   \captionof{table}{Branching fraction of $B_c \to D_s \nu\Bar{\nu}$ decays within SM and NP(MI).}
    \label{tab:bc-decays}
\end{table}

\subsection{$\boldsymbol{V^{\mu}_2(\Bar{3},2,5/6)}$ \textbf{leptoquark:}}
\vspace{0.1cm}

$V^{\mu}_2$ is a doublet vector LQ, where the interaction Lagrangian \cite{Dorsner:2016wpm} in the weak basis and mass basis are, respectively, given by
\begin{align}
\mathcal{L} &\supset  x^{RL}_{ij} \, \bar{d}_R^{c~i}\gamma^\mu V_{2,\mu}^a \, \epsilon^{ab} L_L^{j,b} + x^{LR}_{ij} \, \bar{Q}^{c~i,a}_{L} \gamma^\mu \epsilon^{ab} V_{2,\mu}^b \, e^j_R  + h.c.,
\label{L1}
\end{align}
\begin{align}
\mathcal{L} \supset &- (x^{RL}_{ij}) \bar{d}_R^{c~i}\gamma^\mu V^{1/3}_{2,\mu} \nu_L^j + x^{RL}_{2,ij} \bar{d}_R^{c~i} \gamma^\mu V^{4/3}_{2,\mu} e^j_L  + (V_{CKM}^\dagger x^{LR})_{ij} \bar{u}^{c~i}_L \gamma^\mu V^{1/3}_{2,\mu} \, e^j_R 
\nonumber\\& 
- x^{LR}_{ij} \bar{d}^{c~i}_L \gamma^\mu V^{4/3}_{2,\mu} \, e^j_R  + h.c.
\label{L2}
\end{align}
Within the $V^{\mu}_2$ LQ model,  the following NP WCs can be generated for the $b \to s \ell\bar{\ell}(\nu\bar{\nu})$ transitions :
\begin{itemize}
\item For $b \to s \mu\Bar{\mu}$
\begin{align}
C_9^{\text{NP}} = C_{10}^{\text{NP}} &= 
-\frac{\pi}{\sqrt{2} G_F V_{tb} V_{ts}^* \alpha_{EW}} 
\frac{(x^{LR}_{32})(x^{LR}_{22})^*}{m^2_{(V_2^{\mu~4/3})}}, \\
C_{9}^{' \, \text{NP}} = -C_{10}^{' \, \text{NP}} &= 
-\frac{\pi}{\sqrt{2} G_F V_{tb} V_{ts}^* \alpha_{EW}} 
\frac{(x^{RL}_{32})(x^{RL}_{22})^*}{m^2_{(V_2^{\mu~4/3})}}.
\end{align}
\item For $b \to s \nu\Bar{\nu}$ 

\begin{equation}
\begin{aligned}
    C^{\nu}_R = -\frac{\pi}{\sqrt{2} G_F V_{tb} V_{ts}^* \alpha_{EW}} 
\frac{(x^{RL}_{32})(x^{RL}_{22})^*}{m^2_{(V_2^{\mu~1/3})}} {~\text{~~and}~} \quad &
   C_L^{\nu~NP} = 0.
\end{aligned}
\end{equation}

\end{itemize}
From the $b \to s \mu\Bar{\mu}$ sector, we use the branching fraction $Br(B \to K \mu\Bar{\mu})$ at different $q^2$ bin sizes \cite{CMS:2024syx} to constrain the new couplings. The bins used are [0.1, 0.98], [1.1, 2.0], [2.0, 3.0], [3.0, 4.0], [4.0, 5.0], [5.0, 6.0], [6.0, 7.0], [7.0, 8.0], [11.0, 11.8] and [1.1, 6.0]. 
From the $b \to s \nu\Bar{\nu}$ sector, we consider the measured branching fraction $Br(B \to K \nu\Bar{\nu})$ \cite{Belle-II:2023esi} to constrain the new couplings.

Taking $m_{V^{\mu}_2} = 2$ TeV, we obtain the allowed parametric space for $x^{RL}_{32} \text{~vs~} x^{RL}_{22}$ as shown in Fig. \ref{fig:NP1-NP2}. We  further obtain the best-fit values of the various NP couplings using a $\chi^2$ fitting as given in Table \ref{best-fit}. The obtained $\chi^2/d.o.f.$ is  $6.3116/7 \approx 0.90$.   
\begin{table}[H]
    \centering
    \resizebox{11cm}{!}{%
\begin{tabular}{|c|c|c|c|}
            \hline
             $x^{RL}_{32} = 0.16599 $  & $x^{RL}_{22} = 0.050301$ & $x^{LR}_{32} = 0.092285$ & $x^{LR}_{22} = -0.090918 $ \\ \hline
\end{tabular}
        }
        \end{table}
      \vspace{-32pt}  
       
\begin{table}[H]
    \centering
    \resizebox{9.5cm}{!}{%
\begin{tabular}{|c|c|c|}
            \hline
             $C_9^{\text{NP}} = C_{10}^{\text{NP}} = -1.26$  & $-C_{9}^{' \, \text{NP}} = C_{10}^{' \, \text{NP}} = 1.25$ & $C_R^{\nu} = -1.25 $ \\ \hline
\end{tabular}
        }
        \caption{Obtained best-fit values of various NP couplings.}
        \label{best-fit}
        \end{table}


We also predict the branching fraction of $B_c \to D_s \nu\Bar{\nu}$ decay for different $q^2$ bins within the SM and $V_2^{\mu}$ LQ model using the obtained best-fit values, as given in Table \ref{bs-V2-LHC}. The $q^2$ variation of the differential branching fraction as a function of $q^2$ is shown in Fig. \ref{fig:Bc-to-Ds-V2}. 
 
\begin{table}[ht]
    \centering
    \resizebox{10cm}{!}{%
\begin{tabular}{|c|c|c|c|c|}
            \hline

Bin & $SM$ & $SM$ + $NP$($V_2^{\mu}$)\\
\hline

$0.1 - 0.98 $ &  $(0.435\pm 0.070) \times 10^{-7}$ & $(0.624\pm 0.100) \times 10^{-7}$\\

$1.1 - 2.5 $ &  $(0.699\pm 0.100) \times 10^{-7}$ & $(1.00\pm 0.14) \times 10^{-7}$\\

$2.5 - 4.0 $ &  $(0.761\pm 0.096) \times 10^{-7}$ & $(1.09\pm 0.14) \times 10^{-7}$\\

$4.0 - 6.0 $ &  $(1.04\pm 0.11) \times 10^{-7}$ & $(1.50\pm 0.16) \times 10^{-7}$\\

$1.1 - 6.0 $ &  $(2.50\pm 0.31) \times 10^{-7}$ & $(3.59\pm 0.44) \times 10^{-7}$\\

$0 - q^2_{max} $ &  $(8.70\pm 0.97) \times 10^{-7}$ & $(1.24\pm 0.14) \times 10^{-6}$\\
\hline
\end{tabular}
}
\caption{Branching fraction variation at different $q^2$ bins for $B_c \to D_s \nu\Bar{\nu}$ decay within SM and $V_2^{\mu}$ LQ.}
\label{bs-V2-LHC}
\end{table}

\begin{minipage}{0.45\textwidth}
    \includegraphics[width=\textwidth]{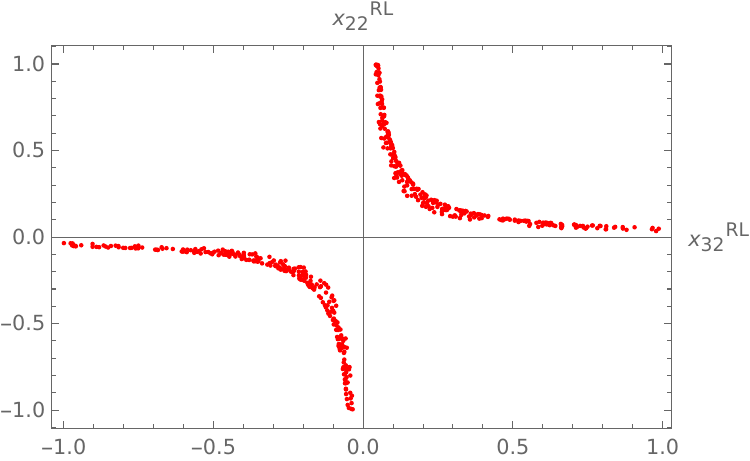} 
    \captionof{figure}{Constraint plot ($x^{RL}_{32} \text{~vs~} x^{RL}_{22}$).}
    \label{fig:NP1-NP2}
\end{minipage}
\hfill
\begin{minipage}{0.45\textwidth}
    \centering
    \includegraphics[width=\textwidth]{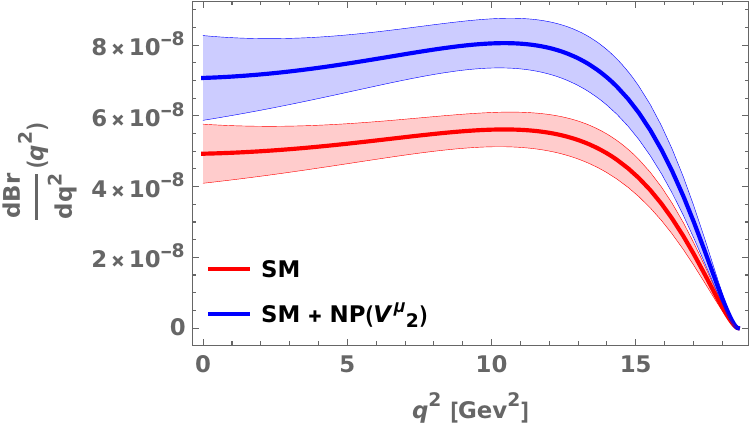} 
    \captionof{figure}{$q^2$-dependency of $dBr(B_c \to D_s \nu\Bar{\nu})$ channel.}
    \label{fig:Bc-to-Ds-V2}
\end{minipage}

\section{Conclusion}
The $B_c \to D_s \nu\Bar{\nu}$ decay mode was analyzed within the SM and beyond the SM - in a model independent approach and $V_2^{\mu}$ leptoquark model. Lattice QCD form factors have been used for the hadronic transition.  Using the new couplings constrained from the $b \to s \ell \bar{\ell}$ and $b \to s \nu \bar{\nu} $ sectors, we predicted the branching fraction of the $B_c \to D_s \nu\Bar{\nu}$ channel. Within the $V_2$ leptoquark  model, the branching fraction was predicted for different $q^2$ bin sizes that are compatible with the LHCb experiment. We found the branching fraction to be sensitive to the new contributions, both in the presence of the $V_2$ leptoquark as well as within the model independent framework. The observations here can be tested and verified with future measurements of the considered mode, such as at the LHCb.   


\section*{Acknowledgements}
Tarun Kumar acknowledges CSIR-HRDG, Government of India for his CSIR-SRF research fellowship.

\bibliographystyle{jhep}
\bibliography{bib}

\end{document}